\documentstyle[twocolumn]{mn}
\input{amssym}
\input{psfig}
\input{epsf.tex}

\newcommand{\mincir}{\raise -2.truept\hbox{\rlap{\hbox{$\sim$}}\raise5.truept
\hbox{$<$}\ }}
\newcommand{\magcir}{\raise -2.truept\hbox{\rla669p{\hbox{$\sim$}}\raise5.truept
\hbox{$>$}\ }}
\newcommand{\minmag}{\raise-2.truept\hbox{\rlap{\hbox{$<$}}\raise 6.truept\hbox
{$>$}\ }}
\newcommand{\be}{\begin{equation}}
\newcommand{\ee}{\end{equation}}
\newcommand{\ba}{\begin{eqnarray}}
\newcommand{\ea}{\end{eqnarray}}
\newcommand{\brr}{\begin{array}}
\newcommand{\err}{\end{array}}
\newcommand{\bc}{\begin{center}}
\newcommand{\ec}{\end{center}}
\newcommand{\etal}{{et al.}~}

\newcommand{\f}{\frac}

\newcommand{\de}{\delta}
\newcommand{\eps}{\epsilon}
\newcommand{\Om}{\Omega}
\newcommand{\fde}{\tilde{\delta}}
\newcommand{\bfx}{{\bf x}}
\newcommand{\bfk}{{\bf k}}
\newcommand{\bfq}{{\bf q}}
\newcommand{\bfu}{{\bf u}}
\newcommand{\bfv}{{\bf v}}
\newcommand{\bfr}{{\bf r}}
\newcommand{\bfS}{{\bf S}}
\newcommand{\bfs}{{\bf s}}
\newcommand{\calB}{{\cal B}}

\newcommand{\calE}{{\cal E}}
\newcommand{\calI}{{\cal I}}
\newcommand{\calL}{{\cal L}}
\newcommand{\calS}{{\cal S}}
\newcommand{\lan}{\langle}
\newcommand{\ran}{\rangle}

\title{Two ways of biasing galaxy formation}

\author[P. Catelan, C. Porciani, \& M. Kamionkowski]{Paolo
Catelan$^1$, Cristiano Porciani$^2$, and Marc Kamionkowski$^1$\\
$^1$ California Institute of Technology, Mail Code 130-33,
Pasadena, CA 91125~~USA\\
$^2$ Racah Institute of Physics, The Hebrew University,
Jerusalem 91904, Israel}

\begin{document}

\maketitle

\begin{abstract}
We calculate the galaxy bispectrum in both real and redshift space
adopting the most common prescriptions for local Eulerian biasing and
Lagrangian evolving-bias model.  We show that the two biasing schemes
make measurably different predictions for these clustering statistics.
The Eulerian prescription implies that the galaxy distribution depends
only on the present-day local mass distribution, while its Lagrangian
counterpart relates the current galaxy distribution to the mass
distribution at an earlier epoch when galaxies first formed. Detailed
measurement of the galaxy bispectrum (of its reduced amplitude) can
help establish whether galaxy positions are determined by the current
mass distribution or an earlier mass distribution. \\

\vspace{1cm}
\end{abstract}

% The different journals have different requirements for keywords.  The
% keywords.apj file, found on aas.org in the pubs/aastex-misc directory, 
% contains a list of keywords used with the ApJ and Letters.  These are 
% usually assigned by the editor, but authors may include them in their 
% manuscripts if they wish. 

%\keywords{galaxies: statistics -- large-scale structure of Universe} 
 
% That's it for the front matter.  On to the main body of the paper.
% We'll only put in tutorial remarks at the beginning of each section
% so you can see entire sections together.

% In the first two sections, you should notice the use of the LaTeX \cite
% command to identify citations.  The citations are tied to the
% reference list via symbolic KEYs.  We have chosen the first three
% characters of the first author's name plus the last two numeral of the
% year of publication.  The corresponding reference has a \bibitem
% command in the reference list below.
%
% Please see the AASTeX manual for a more complete discussion on how to make
% \cite-\bibitem work for you.   

\section{Introduction}
Galaxy clustering in the nearby Universe has been mapped through a
variety of surveys, including different populations of luminous
objects. These run from optical galaxies in the APM, CfA and LCRS
surveys to the sources of the IRAS catalogue at 60 $\mu$m.
Reconstructing the overall mass power spectrum from these data
represents one of the main goals of modern cosmology.  It is already
known, however, that different tracer populations show different
clustering amplitudes even after redshift-space and small-scale
corrections are applied. Thus, their clustering patterns are not
unambiguosly related to any one given mass power spectrum (see Peacock
1999 for a review).

The simplest and most common description of biasing adopted in the
literature is that, at any spatial position $\bfx$, the fluctuation in
the number density of galaxies $\de_g(\bfx)$ responds linearly and
locally to the underlying mass fluctuation $\de(\bfx)$, namely
$\de_g(\bfx) = b^E\de(\bfx)$, where $b^E$ is a space-independent bias
factor (e.g. Dekel \& Rees 1987). As discussed below, higher-order
bias factors can be introduced, but the point is that such a bias
prescription is inherently Eulerian: it relates the present-day galaxy
and mass clustering properties, ignoring their past evolution.
However, if gravity is the main force acting in the Universe, there is
no doubt that galaxy biasing evolves in time, as collapsing mass
fluctuations keep accreting luminous matter onto them, the galaxy
distribution eventually relaxing to the mass one (Fry 1996; Tegmark
and Peebles 1998). So, the biasing in the present-day galaxy
distribution might well be rooted into the deep past of the history of
the Universe: the strong Lyman break galaxy clustering seems to
suggest that this might be the case (Steidel \etal 1997). Any
primordial biasing, arising at the epoch of galaxy formation, cannot
be described by an Eulerian model.  Instead, a Lagrangian one has to
be adopted: it is the primordial fluctuation in galaxies that is
proportional to the mass fluctuation, $\de_g(\bfq) = b^L\de(\bfq)$,
where $\bfq$ denotes the Lagrangian position; in general $b^E$ differs
from $b^L$ and, in principle, higher order factors can be defined.

In this Letter we show that the local Eulerian and Lagrangian bias
models are inconsistent. In fact, the clustering patterns predicted by
the two bias models are different. Specifically we study the galaxy
bispectrum and skewness, both in real and redshift space, on scales
where the mildly non-linear approximation suffices, starting from
Gaussian initial conditions. In Section 2 we review the general
Eulerian and Lagrangian bias models in terms of infinite hierarchies
of bias factors $\{b^E_j\}$ and $\{b^L_j\}$.  In the whole discussion,
these must be considered as free, position-independent, parameters. In
Section 3 we discuss the galaxy bispectrum and skewness in real space,
for both bias models. In Section 4 we carry out the same analysis, but
taking into account the effect of redshift distortions. Section 5
contains our conclusions.
\section{The twofold biasing prescription}
Let us start by fixing the notation of basic quantities.  If
$\varphi_o(\bfq)$ is the primordial gravitational potential (growing
mode only, smoothed on some scale $R_o$ and linearly extrapolated to
the present time), then $\de^{(1)}(\bfq,
z)=D(z)\nabla_q^2\varphi_o(\bfq)$ is the linear density field, and
$D(z)$ its growth factor with $z$ the cosmological redshift [we put
$D(0)=1$].  The linear peculiar velocity is given by
$\bfu^{(1)}(\bfq)= -\nabla_q\varphi_o(\bfq)$, and it is constant in
time. The Eulerian density field will be indicated by $\de(\bfx, z)$
and the $n$-th order perturbative solutions $\de^{(n)}$ are such that
$\de = \sum_n\de^{(n)}$ (Goroff \etal 1986). The Fourier transform is,
e.g., $\fde(\bfk) = \int\!d\bfx\,\de(\bfx)\,\exp{i\bfk\cdot\bfx}$. 
% We assume a generic non-flat cosmology.
%
\subsection{Local Eulerian bias}
In this approach, the galaxy number density field at a given position
$\bfx$ and time $z$ (e.g. `here' and `now') is assumed to be a local
function of the underlying mass density field at the same location and
instant,
$
\de_g(\bfx, z; R) \equiv \calE[\de(\bfx, z; R)],
$
where the smoothing scale $R$ is much larger than the typical size of
the selected objects.  Usually, assuming that $\calE[\de]$ can be
expanded about $\de=0$ as a power series, an infinite set of
``Eulerian bias factors'' $b^E_j$ can be defined (Fry \& Gazta{\~ n}aga 
1993):
\be
\de_g = \sum_{j=0}^\infty \f{b^E_j}{j!}\,\de^{\,j}\;.
\label{2}
\ee
This series is such that $\lan\de_g\ran = 0$ and $\de_g(\de = -1)=
-1$.  The linear coefficient $b_1^E$ corresponds to the usual bias
factor. The origin of this local Eulerian prescription is essentially
phenomenological, and it is {\it a priori} devoid of any insight about
the dynamics of the clustering. Galaxy clustering is analyzed for
instance in terms of $N$-point correlation functions
$\lan\prod_{n=2}^N \de_g(\bfx_n, z)\ran$, and the bias factors are
tuned to fit the observational data.  This is the approach that has
been implicitly adopted in most of the published literature on
biasing, at least in its leading approximation.
\subsection{Local Lagrangian bias}
According to this alternative prescription, the sites of galaxy
formation are identified with specific regions of the primordial
density field.  It is then appropriate to define a ``primordial''
galaxy density field, $\de_g(\bfq)$, measuring the (smoothed)
overdensity of galaxies {\it in fieri} at the Lagrangian position
$\bfq$ at a given time $z$ (formally $z = \infty$, i.e. `there' and
`then') which is biased with respect to the primordial (linear)
density field at the same location and instant, namely
$\de_g(\bfq) \equiv \calL[\eps_o(\bfq)]
=\sum_{j=0}^\infty (b^L_{oj}/j!)\,\eps_o(\bfq)^{\,j};
$
here $\eps_o(\bfq)= \nabla^2\varphi_o(\bfq)$ is the linear density field
extrapolated to the present time. We can equivalently
write, for similarity with the Eulerian case,
\be
\de_g(\bfq) =\sum_{j=0}^\infty \f{b^L_j}{j!}\,\de^{(1)\,j} \;,
\label{4}
\ee
with $\de^{(1)}(\bfq, z)=D(z)\eps_o(\bfq)$ and the Lagrangian factors
$b^L_j$ are defined accordingly in terms of the original $b^L_{oj}$.

Both galaxies and dark matter flow through Eulerian space towards mass
concentrations. Therefore, even assuming that the spots of galaxy
formation can be identified in Lagrangian space, one has to consider
large-scale motions in order to compute the statistics of present-day
structures.  One therefore needs to assign a dynamical prescription,
because evolution changes the original galaxy distribution: this is
the main difference with respect to the Eulerian bias scheme, where no
dynamics is taken into account.

One way is to use galaxies as test particles of the underlying
gravitational field (Fry 1996), then the evolved galaxy density field
at the Eulerian position $\bfx$ and instant $z$ is related to the
primordial galaxy field and evolved density field by the relation
(Catelan \etal 1998)
\be
1 + \de_g(\bfx, z) = [1 + \de_g(\bfq)]\,[1 + \de(\bfx, z)]\;.
\label{5}
\ee
We stress the fact that eq.(\ref{5}) is inherently non-local.
Smoothed regions in Lagrangian space can be mapped to Eulerian space
through the transformation $\bfx = \bfq + \bfS(\bfq, z)$, where $\bfS$
is the displacement vector. In the Zel'dovich (1970) approximation,
the simplest transformation, $\bfS(\bfq, z) = D(z)\bfu^{(1)}(\bfq)$.
Thus, the resulting $\de_g(\bfx,z)$ and $\de(\bfx,z)$ are not
deterministically related. In fact, for any given $\de$ the galaxy
field $\de_g$ can assume different values (see Dekel \& Lahav 1999).
This stochastic behaviour is inherent to the gravitational instability
dynamics.

The question now is the following: are the predictions about the
clustering (in terms of standard statistics as the correlation
functions, for example) as deduced from the local Eulerian and
Lagrangian bias equivalent? In other terms, do there exist two sets of
non-trivial Eulerian and Lagrangian bias factors, $\{b^E_j\}$ and
$\{b^L_j\}$, such that the predictions for galaxy clustering are
identical?  In order to answer to these questions, let us analyze the
galaxy bispectrum from Gaussian initial conditions as induced by
mildly non-linear density evolution.
\section{Galaxy bispectrum}
\subsection{Eulerian bias case}
The lowest order contribution to the galaxy bispectrum 
$
(2\pi)^3\de_D(\bfk_1+\bfk_2+\bfk_3)\,B_g(\bfk_1, \bfk_2, \bfk_3; z)=
\lan\fde_g(\bfk_1, z)\fde_g(\bfk_2, z)\fde_g(\bfk_3, z) \ran 
$
comes from the appearance of non-negligible second-order fluctuations
$\de^{(2)}_g$. From eq.~(\ref{2}), this is
$
\de^{(2)}_g(\bfx, z) = 
b^E_1 \de^{(2)}(\bfx, z) + \f{1}{2}b^E_2 \de^{(1)2}(\bfx, z).
$
(Note that this expression does not have zero mean, so an offset term
should be introduced; however, since we are interested in the spectral
properties of the galaxy clustering, we will ignore it since it
contributes only to $\bfk = {\bf 0}$.) Defining $\nu_{12}\equiv
\bfk_1\cdot\bfk_2/k_1\,k_2$, and the second-order growth factor $E
\approx -\f{3}{7}\Om^{-2/63}D^2$ either in an open Universe with no
cosmological constant (Bouchet \etal 1992) or $E \approx
-\f{3}{7}\Om^{-1/140}D^2$ in a Universe with a cosmological constant
or quintessence (Kamionkowski \& Buchalter 1999), we introduce the
symmetric kernel
\be
J^{(2)}_S \equiv \f{1}{2}\Big(1-\f{E}{D^2}\Big) +
\f{1}{2}\left(\f{k_1}{k_2}+\f{k_2}{k_1}\right)\nu_{12}
+\f{1}{2}\Big(1+\f{E}{D^2}\Big)\,\nu_{12}^2 ,
\label{8}
\ee
and the second-order convolution integral operator $\calI^{(2)} \equiv
\fde^{(1)}\ast\fde^{(1)}$ (Fry 1984).  Finally, we can simply write
$\fde^{(2)} = \calI^{(2)}J_S^{(2)}$ and
\be
\fde^{(2)}_g = \calI^{(2)}\Big( b_1^EJ_S^{(2)} + \f{1}{2}b_2^E \Big)\;. 
\label{10}
\ee
Thus, the galaxy bispectrum is (Matarrese, Verde \& Heavens 1997)
\be
B_g^E = 2 D^4 b_1^{E\,2} \Big[\Big(b_1^E J_S^{(2)} + \f{1}{2}b_2^E\Big) 
P(k_1) P(k_2)\, + \,{\rm c.~t.}\Big],
\label{11}
\ee
where $P(k, z) = D(z)^2\,P(k)$ is the mass linear power
spectrum. The skewness of the galaxy density field smoothed on
scale $R$ is therefore (Fry 1994),
\be
S^E_g(R) = \f{b_1^E(4 - 2\f{E}{D^2}) + 3 b_2^E - b_1^E\gamma(R)}{b_1^{E\,2}}\;,
\label{12}
\ee
where $\gamma= -d\ln\sigma_R^2/d\ln R$ and $\sigma_R^2$ is the rms
density on scale $R$. For a scale-free mass power spectrum $P(k)
\propto k^n$ and a top-hat smoothing function, one obtains $\gamma = n
+ 3$ (Bernardeau 1994). We remind the reader that in the
Einstein--de-Sitter Universe, $4-2E/D^2 = 34/7$.  Mass bispectrum
can be recovered by setting in these formulae $b_1^E=1$ and $b_2^E=0$,
$
B_m^E \equiv 2D^4[J_S^{(2)}P(k_1)P(k_2) +{\rm c.t.}] \equiv B_m\;
$
(Fry 1984).  The growth of $B_m$ is self-similar, i.e. mass particles
do not move from their initial positions, and the wavectors $\bfk$
actually correspond to those positions.
\subsection{Lagrangian bias case}
Let us now repeat the previous calculations assuming the Lagrangian
biasing scheme in eq.(\ref{4}).  In this case, expanding eq.(\ref{5})
up to second-order, we obtain, after Zel'dovich transforming the
Lagrangian coordinate $\bfq$ to the Eulerian one $\bfx$ at $z$,
$\de_g = b_0^L + (1 + b_0^L + b_1^L)\de^{(1)} + \de_g^{(2)}$, where,
\be 
\de^{(2)}_g = (1 + b_0^L)\de^{(2)} -
D\,b_1^L\bfu^{(1)}\cdot\nabla\de^{(1)} + (b_1^L+\f{1}{2}b_2^L)
\de^{(1)2}. 
\label{13b} 
\ee 
This expression generalizes the analogous one in Catelan \etal (1998)
for Press-Schechter dark matter halos, for which $b_0^L = 0$. The
Zel'dovich approximation, adopted here, suffices to transform from
$\bfq$ to $\bfx$: this explains the presence of the inertia term,
proportional to the velocity. (We assume the scale $R_o$ large enough
so that shell-crossing is absent on scale $R$.)  The Fourier transform
of eq.(\ref{13b}) is
\be
\fde^{(2)}_g = \calI^{(2)}\Big((1+b_0^L)J_S^{(2)} + \calB_S^{(2)}\Big)\;, 
\label{14}
\ee
where 
\be
\calB_S^{(2)} \equiv b_1^L + \f{1}{2}b_2^L +
\f{1}{2}b_1^L\,\Big(\f{k_1}{k_2}+\f{k_2}{k_1}\Big)\nu_{12}
\label{14b}
\ee
describes the effects of Lagrangian biasing during the mildly
nonlinear regime. So, the galaxy bispectrum is
\begin{eqnarray}
B_g^L& =& 2 D^4 (1 + b_0^L + b_1^L)^2\Big[\Big((1+b_0^L)J_S^{(2)} + 
\calB_S^{(2)}\Big) \nonumber \\
&\times& P(k_1)\,P(k_2) + {\rm c.~t.}\Big],
\label{15}
\end{eqnarray}
and the galaxy skewness turns out to be
\be
S_g^L(R)\!=\!\f{(1\!+\!b_0^L)(4\!-\!2\f{E}{D^2})\!+\!6b_1^L\!+\!3b_2^L
\!-\!(1\!+\!b_0^L\!+\!b_1^L)\gamma(R)}{(1 + b_0^L + b_1^L)^2}.
\label{16}
\ee
We emphasize the fact that it is the term
$\f{1}{2}b_1^L\,(k_1/k_2+k_2/k_1)\nu_{12}$ in $\calB_S^{(2)}$, Fourier
transform of the inertia term $-b_1^L\bfu^{(1)}\cdot\nabla\de^{(1)}$
in eq.(\ref{13b}), which carries the signature of the gravitational
dynamics, inherently absent in the Eulerian description. Clearly, such
a signature would reflect into a distinctive shape dependence of the
galaxy bispectrum, best quantified by the `effective' amplitude $Q$
(see below).
\subsection{Disentangling the two biasing schemes}
\subsubsection{Galaxy Bispectrum and Skewness}

The different clustering predictions of the two biasing models may be
emphasized simply by calculating the difference $\Delta B_g$ of the
bispectra in eq.(\ref{11}) and eq.(\ref{15}) or of the skewnesses
$\Delta S_g$ in eq.(\ref{12}) and eq.(\ref{16}). It can be easily
verified that no two sets of nontrivial independent bias factors
$\{b_j^E\}$ and $\{b_j^L\}$ can be found such that $\Delta B_g = 0
=\Delta S_g$.  We can explicitly write the final expressions assuming
that at least the lowest-order bias factors are related, namely $b_1^E
= 1+ b_0^L + b_1^L$. This last relation may be easily derived in
linear regime, but it shows to be preserved even during the mildly
non-linear regime (Mo \& White 1996; Mo, Jing \& White 1997 for the
case $b_0^L=0$). Thus $\Delta B_g = B^E_g - B^L_g$ is
\begin{eqnarray}
\Delta B_g \! & \!=\!\!&\! \!D^4\Big(1 + b_0^L + b_1^L\Big)^2
\Big\{\Big[(b_2^E - b_2^L) - b_1^L \Big(1+\f{E}{D^2}\Big)\nonumber \\
\!&\! \times \!&\! 
(1-\nu_{12}^2)\Big]\, P(k_1)\,P(k_2) + {\rm c. ~ t.} \Big\} \nonumber \\
&\equiv& \Delta B_g(1,2)+\Delta B_g(1,3) + \Delta B_g(2,3)\;.
\label{17}
\end{eqnarray}
Correspondingly, the skewness difference is
\be
\Delta S_g = \f{3(b_2^E - b_2^L) - 2b_1^L (1 + \f{E}{D^2})}{(1 +
b_0^L + b_1^L)^2}\;.
\label{18}
\ee
Intriguingly, the dependence on the filtering scale cancels out, and
we are left with a residual difference in the skewness of the two bias
models that is scale-independent. Expressions for the
Einstein-de-Sitter Universe can be recovered by setting $1+E/D^2 =
4/7$.

The two biasing schemes cannot be distinguished by measuring the
skewness alone, but they can be distinguished from the shape
dependence of the bispectrum.  The important point is that
$\calB_S^{(2)}$, the new term that arises in the Lagrangian-biasing
bispectrum [cf. eq. (11)] is linearly independent of the mass kernel
$J_S^{(2)}$ and the constant $b_2^E/2$, the two terms that make up the
Eulerian-biasing bispectrum [cf. eq. (6)].  Thus, {\it no combination
of parameters $b_1^E$ and $b_2^E$ can allow the Eulerian-biasing
bispectrum to mimic the Lagrangian-biasing bispectrum}.

Since the bias of high-redshift populations is $\gtrsim$ 3 while that
of local populations are closer to unity, we heuristically expect bias
evolution of order unity, and thus the Lagrangian-bias parameters
$b_0^L$ and $b_1^L$ to be quantities of order unity (further modeling
is required to give a more precise answer).  It is likely that surveys
such as the SDSS and 2dF will be able to measure the bispectrum with
enough precision to distinguish the predictions of Eulerian and
Lagrangian biasing if the bias parameters are of order unity.
Matarrese, Verde \& Heavens (1997) have determined that SDSS/2dF data
should be able to determine $b_1^E$ and $b_2^E$ to roughly a few
percent within the context of Eulerian-bias models; that is, the
coefficients of $J_S^{(2)}$ and the constant in eq. (6) can be
determined to a few percent.  To distinguish between Eulerian and
Lagrangian biasing requires that the data be fit to an additional
term, $\calB_S^{(2)}$, as well.  Although we have not revisited the
calculation in detail, it seems reasonable that if the coefficients of
$J_S^{(2)}$ and the constant can be fit to a few percent, then the
coefficients of $\calB_S^{(2)}$ can be fit with a precision not much
poorer.  In this case, the data can discriminate between Eulerian- and
Lagrangian-biasing schemes.
\subsubsection{$Q$-Amplitudes}
Eulerian and Lagrangian biasings predict for the galaxy
bispectrum two different shape dependences, which cannot be obtained
from one another by simply tuning the bias parameters $b_1^L, b_2^L$
and $b_1^E, b_2^E$. An efficacious way of emphasizing such a shape
dependence is through the bispectrum `amplitudes' $Q^E_g$ and $Q_g^L$,
where, for example,
\be
Q_g^L \equiv \f{B_g^L(\bfk_1, \bfk_2, \bfk_3, z)}
{[P_g^L(k_1, z)P_g^L(k_2, z)+ {\rm c.t.}]}\;,
\label{20b}
\ee
and similarly for $Q_g^E$ (Fry 1984). $Q$-amplitudes are essentially
insensitive to the scale and the overall geometry. We obtain 
\be
Q_g^E = Q_g^L + \f{\Delta B_g}{\sum P_g P_g}\;.
\label{20c}
\ee
It is very useful to express $Q_g^E$ and $Q_g^L$ in terms of the
bispectrum amplitude of the underlying mass density distribution
$Q_m\equiv B_m/\sum P P$.  We have, respectively,
\be
Q_g^E = \f{Q_m}{b_1^E} + \f{b_2^E}{b_1^{E\,2}}\;, 
\label{20d}
\ee
for the Eulerian amplitude (Fry 1994), and, for the Lagrangian
amplitude,
\ba
Q_g^L &=& \f{Q_m}{1 + b_0^L + b_1^L} + \f{b_2^L}{(1+b_0^L + b_1^L)^2}
\nonumber \\
&+& \f{(1+E/D^2)\,b_1^L}{(1+b_0^L + b_1^L)^2}\,\f{[(1-\nu_{12}^2)P(k_1)P(k_2)+
{\rm c.t.}]}
{[P(k_1)P(k_2) + {\rm c.t.}]}\;.
\label{20e}
\ea
The novelty contained in eq.(\ref{20e}) is the angular dependence
appearing in the right hand side: it is not related to the
$Q_m$--shape dependence, and it is independent from the values of
$\{b_i^L\}$. Measurements of $Q_g$ from galaxy catalogs for two
different shapes of the triangle $\bfk_1 + \bfk_2 + \bfk_3 = {\bf 0}$
can in principle disentangle the two biasing factors $b_1$ and $b_2$
and the two biasing schemes as well. In Figure 1 we plot $Q_g^E$ and
$Q_g^L$ for a $\Lambda CDM$ model; the values of the biasing factors
$b_{1,2}^E$ and the choice of the scales are based on Scoccimarro et
al. (2000).

More sophisticated and predictive relations may be proposed if one
assumes that the set of Lagrangian bias factors $\{b_j^L\}$ are not
free parameters, as in the present discussion, but rigourously
computed within the framework of a given theoretical model. The
`excursion set' formalism (Peacock \& Heavens 1990; Bond \etal 1991),
for example, where dark-matter halos are identified by
first-upcrossings of a collapse thereshold, predicts that $b_1^L$ and
$b_2^L$ are functions of both halo size and redshift (Mo \& White
1996; Porciani \etal 1998).
\section{Redshift distortion effects}
Given that the two biasing schemes are in principle distinguishable,
we proceed to calculate the Eulerian and Lagrangian bispectra in
redshift space, which is where they are most likely to be measured.
Peculiar motions associated with structures on any scale distort the
clustering pattern in redshift space (Kaiser 1987). So, in order to
reconstruct the actual distribution of galaxies from redshift
catalogues, we must be able to invert the distortion process. This can
be easily done if we consider a distant region of the Universe so that
the distortions essentially occur along the line-of-sight, and we
restrict to large scales for which the mildly non-linear approximation
suffices. If $\bfr$ is the physical coordinate, and
$u=\bfv\cdot\bfr/r$ is the line-of-sight component of the peculiar
velocity $\bfv$, assuming that the observer's peculiar velocity is
zero, the apparent galaxy fluctuation $\de_g^s(\bfs)$ at the apparent
position $\bfs = (1 + u/r)\bfr$ is related to the actual one $\de_g$
computed at the same apparent position by the relation
\be
\de_g^s(\bfs) = \de_g(\bfs) - u'(\bfs) - \Big[u(\bfs)
\Big(\de_g(\bfs) - u'(\bfs)\Big)\Big]'\;.
\label{23}
\ee
Here $u'$ indicates the first radial derivative of $u$. Since in this
section we will compute the effects of redshift distortions on the
galaxy bispectrum, both for a local Eulerian and Lagrangian bias, only
corrections up-to second order are considered. We remind the reader
that in the distant-observer limit, the Fourier transform of $d/dr
\rightarrow ik\mu$ where $\mu = \bfk\cdot \bfr/kr$ and ${\tilde u} =
i\mu f(\Om){\tilde \eta}/k$, where $f(\Om)\approx \Om^{0.6}$, $a$ is
the universal scale factor, and $\eta$ is the divergence of the
velocity field.
\begin{figure}
%\setlength{\unitlength}{1cm}
%\centering
\centerline{
\epsfxsize= 8 cm \epsfbox{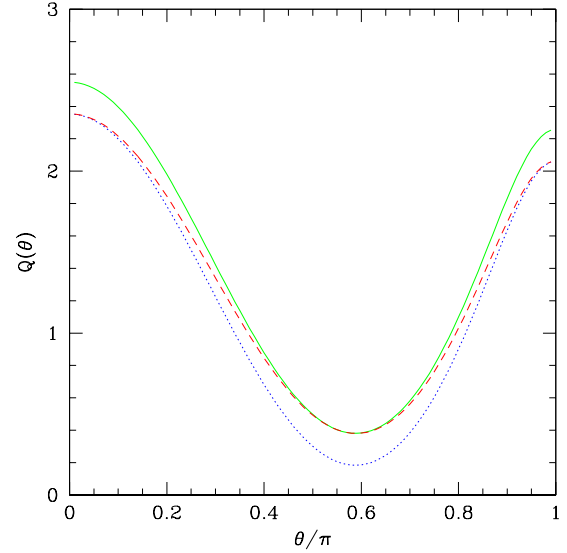}
}
%\begin{picture}(8,10)
%\put(0.0,0.0){\special{psfile=l.ps vscale=60 hscale=60
%angle=0 voffset=-100 hoffset=-80}}
%\end{picture}
\caption{The halo bispectrum amplitude $Q$ for configurations with
sides $k_1 = 0.05\,h_{65}/{\rm Mpc}$ and $k_2 = 0.1\,h_{65}/{\rm Mpc}$
separated by an angle $\theta$ for a linear $\Lambda-$CDM power
spectrum ($\Omega_m=0.3$, $\Omega_\Lambda=0.7$, $n=1$, $\sigma_8=0.9$).
The prediction of the local Eulerian bias model with the bias
parameters estimated from the {\it IRAS} QDOT 2 Jy redshift catalogue
($b_1^E=0.76$, $b_2^E=-0.33$; Scoccimarro \etal 2000) is represented by
a dashed line.  The continuos and dotted lines represents two local
Lagrangian bias models with $ b_0^L=0$ and $b_1^L=b_1^E-1=-0.24$.  The
value of $b_2^L$ has been fixed to match the Eulerian bias prediction
for $Q$ at its minimum and maximum value (respectively, $b_2^L=-0.39$
and $b_2^L=-0.58$): for these specific configuration and choice of
parameters, discrepancies between the predictions of the two biasing
schemes are about $10 \%$ for the $Q$-tails and about 50\% for the
$Q$-trough.}
\label{fig}
\end{figure}
\subsection{EB galaxy bispectrum in redshift space}
In this case, inserting eq.(\ref{10}) into eq.(\ref{23}), the Fourier
transform of $\de_g^s(\bfs)$ is given by
\be
\fde_g^s = (b_1^E + \mu^2\,f)\fde^{(1)} + \calI^{(2)}\calS_E^{(2)}\;,
\label{24}
\ee
where the redshift-distorted symmetric Eulerian-bias kernel is
\begin{eqnarray}
\calS_E^{(2)}
&\equiv& 
b_1^E\,J_S^{(2)}         + 
\mu^2\,f\,K_S^{(2)}      + 
\f{1}{2}b^E_2           \nonumber \\
&+&\f{1}{2}\,b_1^E\,f\Big[
\mu_1^2 + \mu_2^2 +
\mu_1\,\mu_2
\Big(\f{k_1}{k_2} + 
\f{k_2}{k_1}\Big) 
\Big] \nonumber \\
&+& f^2\Big[
\mu_1^2\,\mu_2^2 +
\f{1}{2}\mu_1\,\mu_2
\Big(
\mu_1^2\f{k_1}{k_2} + 
\mu_2^2\f{k_2}{k_1}\Big) 
\Big]   \;.                   
\label{25}
\end{eqnarray}
The quantity $K_S^{(2)}$ describes the second-order contribution to
$\eta$ (Goroff \etal 1986). The distorted galaxy bispectrum is
(Heavens, Matarrese \& Verde 1998)
\be
B_g^s(E)=
2D^4(b_1^E+\mu_1^2f)(b_1^E+\mu_2^2f)
\calS_E^{(2)}P(k_1)P(k_2)+{\rm c.t.}
\label{26}
\ee
\subsection{LB galaxy bispectrum in redshift space}
We adopt in this case the expression in eq.(\ref{14}), obtaining, after
analogous calculations, 
\be
\fde_g^s = (1 + b_0^L + b_1^L + \mu^2f)\fde^{(1)} + \calI^{(2)}\calS_L^{(2)}\;.
\label{27}
\ee
Thus, the galaxy bispectrum is
\begin{eqnarray}
B_g^s(L)\! &\!=\!&\! 2\,D^4\,(1+b_0^L+b_1^L+\mu_1^2\,f)\,
(1+ b_0^L + b_1^L+\mu_2^2\,f) \nonumber \\         
\!&\!\times \!&\!\calS_L^{(2)}\,P(k_1)\,P(k_2)+{\rm c.~t.}\;,
\label{28}
\end{eqnarray}
where, in this bias prescription, the redshift-distorted second-order
Lagrangian-bias kernel $\calS_L^{(2)}$ is
\begin{eqnarray}
\calS_L^{(2)} &\equiv &
(1+b_0^L)J_S^{(2)} + \calB_S^{(2)} +
\mu^2\,f\,K_S^{(2)}  \nonumber \\     
&+&\f{1}{2}\,(1 + b_0^L + b_1^L)\,f
\Big[\mu_1^2 + \mu_2^2 +
\mu_1\,\mu_2
\Big(\f{k_1}{k_2} + 
\f{k_2}{k_1}\Big)\Big] \nonumber \\
&+&f^2\Big[
\mu_1^2\,\mu_2^2 +
\f{1}{2}\mu_1\,\mu_2
\Big(
\mu_1^2\f{k_1}{k_2} + 
\mu_2^2\f{k_2}{k_1}\Big) 
\Big]   \;.                   
\label{29}
\end{eqnarray}
\subsection{Comparing bias in redshift space}
If we assume, once again, the validity of the algebric relation $b_1^E
= 1 + b_0^L + b_1^L$, it follows that between the redshift-distorted
kernels holds the relation
$
\calS_E^{(2)} = \calS_L^{(2)} + b_1^L\,J_S^{(2)} +\f{1}{2}b_2^E
-\calB_S^{(2)}. 
$
This relation should be immediately compared with the one in
eq.(\ref{17}), to understand that we obtain the concise expression
between the quantities $\Delta B^s_g$ and $\Delta B_g$ which emphasize 
the inconsistency between the two biasing prescriptions,
\be
\Delta B^s_g = (1 + \mu_1^2 \beta)\, (1 + \mu_2^2 \beta) \,\Delta B_g(1,2) + 
{\rm c.~t.}\;,
\label{31}
\ee
where $\beta \equiv f/(1 + b_0^L + b_1^L)$. Thus, the only redshift
effect on the quantity $\Delta B_g(i,j)$ comes from the first-order
distortion of the galaxy number density field, $\fde_g^{s(1)} = (1 +
\mu^2\,\beta)\fde_g^{(1)}$. It has to be like that, if one thinks that
the distortion effects due to peculiar motions are either independent
of the bias factors or proportional to the first-order bias factors,
then they cancel out. In redshift space, eq.(\ref{20c}) becomes
\be
Q_g^s(E) = Q_g^s(L) + \f{\Delta B_g^s}{\sum P_g^s P_g^s}\;.
\label{31b}
\ee
Though the structure of the expression (\ref{20c}) is preserved in
redshift space, $Q_g(Q_m)$-relations like those in eqs.(\ref{20d}) and
(\ref{20e}) are not.  A comprehensive investigation of the effects of
redshift distortions on $B_g^E$ and $B_g^L$ is in progress; see also 
Scoccimarro et al. (1999) for an analysis of $B_g^s(E)$.
\section{Discussion and conclusions}
We compared the galaxy clustering predictions of the local Eulerian
bias scheme versus those of the Lagrangian one. We showed that the two
bias models are inconsistent, since the predicted three-point galaxy
correlations are different. A similar inconsistency certainly
characterizes correlations of higher order, or of lower order but
higher perturbative corrections. Qualitatively, these results are
independent on whether the Lagrangian zero-order bias factor $b_0^L$
is zero, as for Press-Schechter dark matter halos, or not, as in the
most general case we have considered here.  The galaxy bispectrum is
much better suited to distinguish between the two bias models than the
corresponding skewness, since the latter is spatially averaged: the
bispectrum depends on the shape of the triangle $\bfk_1 + \bfk_2
+\bfk_3 = {\bf 0}$, thus two shapes can disentagle the two bias
factors $b_1$ and $b_2$ (Matarrese, Verde \& Heavens 1997; Scoccimarro
2000) and the two bias models. The shape dependence is best
quantified by the $Q$-amplitudes discussed in Subsection 3.3.: the
reduced amplitude of the bispectrum from Lagrangian bias in
eq.(\ref{20e}) displays a dependence on the triangle configuration
which is not contained in eq.(\ref{20d}), and, since it is dynamically
induced, which is independent from the bias factors.
The next generation redshift catalogues, as the ongoing Two Degree
Field Survey and the Sloan Digital Sky Survey, will contain enough
galaxies to establish whether $B_g^s(E)$ in eq.(\ref{26}) or
$B_g^s(L)$ in eq.(\ref{28}) better fits the observational data, but
they cannot be both correct, whatever the assumed cosmology.

Both bias schemes represent rather extreme and idealized approaches.
Lagrangian models imply a sort of infinite-memory process, since the
sites for galaxy formation are known from the beginning, and dynamical
evolution changes their spatial distribution.  On the other hand, in
local Eulerian schemes galaxies are simply `painted' on a snapshot of
the density field, without a record of the past. However, even though
real galaxy formation is probably a process with intermediate
characteristics with respect to the biasing schemes discussed here,
recent models based on a Lagrangian selection of the sites for object
formation were shown to be very successful in reproducing the
clustering of dark-matter halos found in numerical simulations
(e.g. Catelan, Matarrese \& Porciani 1998; Porciani, Catelan \& Lacey
1999).  The issue discussed in this Letter surely deserves further
investigation, both in real and redshift space. It would be of
interest to test which biasing scheme better describes galaxy power
spectrum and bispectrum from a combination of numerical simulations
and semianalytic models (Porciani et al., in preparation). \\

\noindent {\bf Acknowledgments} We thank Sabino Matarrese for
discussions. The anonymous referee improved the presentation of these
results. CP is supported by a Golda Meir Fellowship.  This work was
supported at Caltech in part by the DoE, NSF, and NASA.

\end{document}